\documentclass[12pt,preprint]{aastex}
\newcommand{\kms}{${\rm km~s}^{-1}$}

\begin{document}

\title{Two Distant Halo Velocity Groups Discovered by the Palomar Transient Factory}

\author{
Branimir Sesar\altaffilmark{\ref{Caltech},\ref{email}},
Judith G.~Cohen\altaffilmark{\ref{Caltech}},
David Levitan\altaffilmark{\ref{Caltech}},
Carl J.~Grillmair\altaffilmark{\ref{Spitzer}},
Mario Juri\'c\altaffilmark{\ref{LSST},\ref{Steward}},
Evan N.~Kirby\altaffilmark{\ref{Caltech},\ref{Hubble}},
Russ R.~Laher\altaffilmark{\ref{Spitzer}},
Eran O.~Ofek\altaffilmark{\ref{Weizmann}},
Jason A.~Surace\altaffilmark{\ref{Spitzer}},
Shrinivas R.~Kulkarni\altaffilmark{\ref{Caltech}},
Thomas A.~Prince\altaffilmark{\ref{Caltech}}
}
\altaffiltext{1}{Division of Physics, Mathematics and Astronomy, California
                 Institute of Technology, Pasadena, CA 91125,
                 USA\label{Caltech}}
\altaffiltext{2}{Spitzer Science Center, California Institute of Technology,
                 Pasadena, CA 91125, USA\label{Spitzer}}
\altaffiltext{3}{Large Synoptic Survey Telescope Corp., 933 North Cherry
                 Avenue, Tucson, AZ 85721, USA\label{LSST}}
\altaffiltext{4}{Steward Observatory, University of Arizona, Tucson, AZ 85721,
                 USA\label{Steward}}
\altaffiltext{5}{Hubble Fellow\label{Hubble}}
\altaffiltext{6}{Benoziyo Center for Astrophysics, Weizmann Institute of
                 Science, 76100 Rehovot, Israel\label{Weizmann}}
\altaffiltext{7}{Corresponding author: bsesar@astro.caltech.edu\label{email}}

\begin{abstract}
We report the discovery of two new halo velocity groups (Cancer groups A and B)
traced by 8 distant RR Lyrae stars and observed by the Palomar Transient
Factory (PTF) survey at ${\rm R.A}\sim129\arcdeg$, ${\rm Dec}\sim20\arcdeg$
(${\rm l}\sim205\arcdeg$, ${\rm b}\sim32\arcdeg$). Located at 92 kpc from the
Galactic center (86 kpc from the Sun), these are some of the most distant
substructures in the Galactic halo known to date. Follow-up spectroscopic
observations with the Palomar Observatory 5.1-m Hale telescope and W. M. Keck
Observatory 10-m Keck I telescope indicate that the two groups are moving away
from the Galaxy at $\bar{v}^A_{gsr} = 78.0\pm5.6$ {\kms} (Cancer group A) and
$\bar{v}^B_{gsr} = 16.3\pm7.1$ {\kms} (Cancer group B). The groups have velocity
dispersions of $\sigma^A_{v_{gsr}}=12.4\pm5.0$ {\kms} and
$\sigma^B_{v_{gsr}}=14.9\pm6.2$ {\kms}, and are spatially extended (about
several kpc) making it very unlikely that they are bound systems, and are more
likely to be debris of tidally disrupted dwarf galaxies or globular clusters.
Both groups are metal-poor (median metallicities of ${\rm [Fe/H]^A = -1.6}$
dex and ${\rm [Fe/H]^B=-2.1}$ dex), and have a somewhat uncertain (due to small
sample size) metallicity dispersion of $\sim0.4$ dex, suggesting dwarf galaxies
as progenitors. Two additional RR Lyrae stars with velocities consistent with
those of the Cancer groups have been observed $\sim25\arcdeg$ east, suggesting
possible extension of the groups in that direction.
\end{abstract}


\section{Introduction\label{introduction}}

In the last decade, deep and wide-area astronomical surveys, such as the Sloan
Digital Sky Survey (SDSS; \citealt{yor00}) and Two-Micron All Sky Survey,
(2MASS; \citealt{skr06}) have enabled detection of substructures in the halo
both as stellar overdensities in space and as moving groups. The tidal streams
of the disrupting Sagittarius dwarf galaxy \citep{igi94} are the best example
of such substructures, with streams wrapping around most of the sky
\citep{maj03}. Other known substructures include the Pisces Overdensity
(``Clump J'' in \citealt{ses07}; \citealt{wat09}; \citealt{ses10a}), the GD$-$1
stream \citep{gd06}, the Orphan stream \citep{gri06, bel07}, and several other
overdensities and streams\footnote{A fairly extensive list of halo streams
and overdensities can be found at \url{http://homepages.rpi.edu/~newbeh/mwstructure/MilkyWaySpheroidSubstructure.html.}}.

While many halo substructures have been discovered so far, the motivation for
finding and characterizing more of them remains strong. For example, at larger
galactocentric distances ($R_{\rm GC}\gtrsim15$ kpc), the Galactic
gravitational potential is dominated by dark matter and observations of more
distant tidal streams can be used to constrain the shape, orientation, and mass
of the Milky Way's dark matter halo (e.g., \citealt{lm10}, \citealt{wat10}).
Other motivations for completing the census of halo substructures are to better 
constrain the severity of the so-called ``missing satellite problem''
\citep{kly99, moo99} and the contribution of the accretion of dwarf satellite
galaxies to the formation of the Galactic halo.

Finding distant halo substructures, however, is not an easy task. One approach
to this problem is to look for spatial groups of (non-variable) sources that
are consistent in color-magnitude space with, for example, an old population of
stars at a fixed distance (e.g., \citealt{wil05, gri06, bel07}). While this
technique has been employed quite successfully in the past, its strongest
limitation when looking for distant halo substructures using ground-based
imaging data is the separation of stars from galaxies at faint magnitudes. At
magnitudes fainter than $r\sim21$ (corresponding to a heliocentric distance of
$\sim30$ kpc for metal-poor main-sequence turn-off stars), the number of field
Milky Way stars per unit magnitude decreases with increasing magnitude (i.e.,
the number density profile of Galactic halo stars steepens with distance from
the Galactic center from a power-law with index $n=-2.6$ to that with
$n\lesssim-3.8$; \citealt{sji11, dbe11}), while the number of galaxies per unit
magnitude increases. In addition, morphological separation of stars and
galaxies (e.g., SDSS resolved-unresolved source classification;
\citealt{lup02}) becomes increasingly unreliable towards faint magnitudes as
the signal-to-noise ratio decreases. The increasing contribution of galaxies in
point-source catalogs at faint magnitudes causes a sharp increase in the noise
from which the signal of a true low surface brightness stellar system must be
detected.

Instead of looking for clustering in samples of non-variable tracers, which may
get contaminated by galaxies at faint magnitudes, we should look for clustering
in samples of variable tracers with distinctive light curves (e.g., RR Lyrae
stars) that are difficult to confuse with galaxies. RR Lyrae stars have several
advantages when used to map the Galactic halo because they i) are old, evolved
stars and therefore trace old stellar populations \citep{smi95}; ii) have
distinct light curves which make them easy to identify given multi-epoch
observations (peak-to-peak amplitudes of $V\sim1$ mag and periods of $\sim 0.6$
days); and iii) are bright, standard candles ($M_V=0.6$ mag at
${\rm [Fe/H]}=-1.5$ dex, with $\sim7\%$ uncertainty in distance) that can be
detected at large distances (5-120 kpc for $14 < V < 21$). The steepening of
the stellar density profile beyond 30 kpc is actually beneficial for searches
that employ RR Lyrae stars because it reduces the pool of stars that can form
false spatial groups, and therefore increases the contrast between stars
associated with halo substructures and stars associated with the smooth halo
(e.g., the Pisces Overdensity shown in Figure 11 of \citealt{ses10a}). Due to
the steepening of the stellar density profile beyond 30 kpc, groups of distant 
RR Lyrae stars are more likely to be real halo substructures than chance
associations of stars. While the probability of chance association at large
distances and small angular scales is very small, it is not zero and
spectroscopic followup is still needed to test whether the stars in a spatial
group also form a moving group.

In this paper we use the line of reasoning presented above, and successfully
used by \citet{kol09} and \citet{ses10b}, as a motivation to study a group of 8
RR Lyrae stars found during a preliminary search for halo substructures in
regions of the sky observed by the Palomar Transient Factory
survey\footnote{\url{http://www.astro.caltech.edu/ptf}}. The data set and the
algorithm used to select RR Lyrae stars are described in
Sections~\ref{database} and~\ref{selection}. The spectroscopic observations,
their reduction, and measurement of line-of-sight velocities and metallicities 
are described in Section~\ref{spectro_obs}. The results are discussed in
Section~\ref{discussion} and our conclusions are presented in
Section~\ref{conclusions}.

\section{Overview of the Palomar Transient Factory Survey\label{database}}

The Palomar Transient Factory (PTF; \citealt{law09}, \citealt{rau09}) is a
synoptic survey designed to explore the transient sky. The project utilizes the
48-inch Samuel Oschin Schmidt Telescope on Mount Palomar. The telescope has a
digital camera equipped with 12 CCDs (one of which is not active), each
2K$\times$4K pixels \citep{rah08}. Each PTF image covers 7.26 deg$^2$ with a
pixel scale of $1.01\arcsec$. By the end of 2011, the PTF observed $\sim8000$
deg$^2$ of sky at least 30 times in the Mould-$R$ filter\footnote{The PTF
Mould-$R$ filter is similar in shape to the SDSS $r$-band filter, but shifted
{27 \AA} redward.} (hereafter,the $R$-band filter) and $\sim1800$ deg$^2$ in the
SDSS $g^\prime$ filter.

Some regions of the sky, known as the ``high-cadence'' regions in PTF, have
been observed more than 100 times. One such region on which we focus in this
paper is a 45 deg$^2$ field roughly centered on the Praesepe\footnote{Also
known as the Beehive Cluster and M44.} cluster (08$^h$ $40\arcmin$ $24\arcsec$,
+19$\arcdeg$ 41$\arcmin$), which has been imaged more than 200 times by PTF in
the $R$-band \citep{agu11, pol12}. Relative to other, less observed areas, the
Praesepe region is interesting because, due to chance and a large number of
observations, it has a non-negligible number of images that are deeper than the
nominal $5\sigma$ faint limit at $R=20.6$ (i.e., images that were taken in
better than average conditions). In turn, these deeper images increase the
likelihood of finding distant RR Lyrae stars and halo substructures, if any are
present.

We use PTF data processed by the photometric pipeline, which is hosted by the
Infrared Processing and Analysis Center (IPAC). This pipeline performs final
image reduction, source extraction, and photometric and astrometric calibration
(\citealt{gri10, ofe12}; Laher, R. et al. 2012, in prep). The photometric
uncertainty provided by this pipeline is smaller than $\sim0.01$ mag for $R<16$
sources and increases to 0.2 mag at $R=20.6$. The algorithm used for
photometric calibration is based on the one presented in \citet{hon92} and
modifications in \citet{ofe11} and \citet{lev11}. Relative to the reference
UCAC-3 astrometric catalog \citep{zac10}, the astrometric precision of PTF
coordinates is about $0.1\arcsec$ in right ascension and declination. To enable
fast processing and manipulation, the data used in this paper were stored in
the Large Survey Database
format \citep[LSD\footnote{\url{http://lsddb.org}};][]{lsd}.

\section{Selection of RR Lyrae Stars\label{selection}}

\begin{figure}
\epsscale{0.9}
\plotone{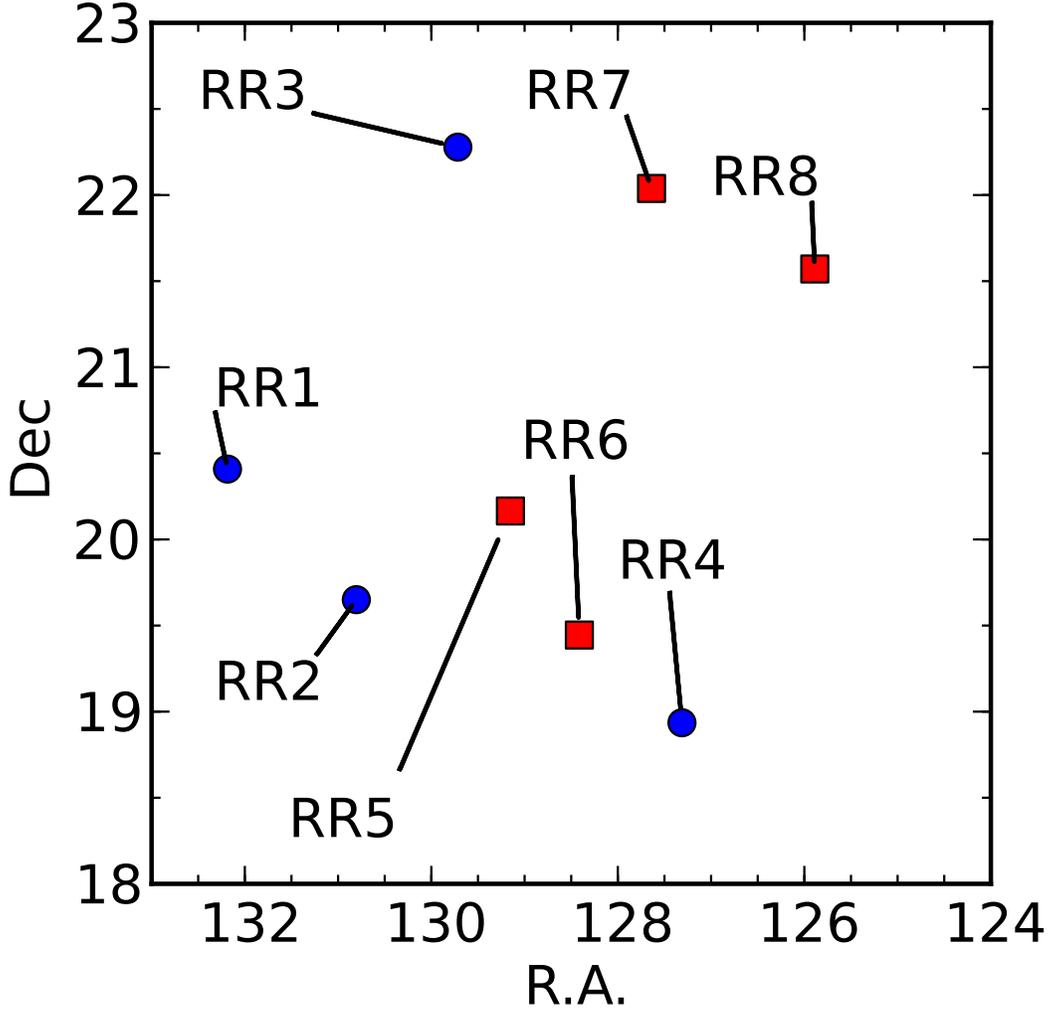}
\caption{
Spatial distribution of 8 $ab$-type RR Lyrae stars studied in this work, which
are located at distances of 77-96 kpc from the Sun (84-102 kpc from the Galactic
Center). Their positions and light-curve parameters are listed in
Table~\ref{table-positions}. The symbols indicate the halo velocity group
membership for each of these stars in either the $\bar{v}_{gsr}\sim 80$ km
$s^{-1}$ group (blue circles) or the $\bar{v}_{gsr}\sim 15$ km $s^{-1}$ group
(red squares, see Section~\ref{discussion}). The lines simply connect labels
and symbols. According to the completeness analysis presented in
Section~\ref{completeness}, these are the only RRab stars in this region of the
Galactic halo.
\label{fig1}} 
\end{figure}

In Figure~\ref{fig1} we show the spatial distribution of 8 distant
(galactocentric distance $R_{GC}>84$ kpc) $ab$-type RR Lyrae (RRab) stars
observed by PTF in the Praesepe region. These stars were selected during a
preliminary search for RR Lyrae stars in the available PTF observations. The
selection procedure we used is similar to the one outlined by \citet{ses10a}
and is briefly described below. The completeness of this sample is estimated in
Section~\ref{completeness}.

First, we searched for all PTF sources positionally associated within
$1.5\arcsec$ of point-like sources in the SDSS Data Release 8 catalog
\citep{aih11} that have single-epoch SDSS colors consistent with colors of RR
Lyrae stars (Equations 6 to 9 from \citealt{ses10a}):
\begin{eqnarray}
0.75 < u-g < 1.45 \\
-0.25 < g-r < 0.4 \\
-0.2 < r-i < 0.2 \\
-0.3 < i-z < 0.3\label{single_epoch_colors}.
\end{eqnarray}
In the above equations, SDSS colors have been corrected for ISM extinction
using the dust map created by \citet{SFD98}.

Variable sources were selected from this color-selected sample by requiring
that PTF $R$-band light curves have root-mean-square (rms) scatter greater than
0.1 mag and $\chi^2$ per degree of freedom greater than 3 (calculated with
respect to the mean magnitude associated with a given source). The five most
likely periods of variability were obtained for each variable PTF source by
running an implementation of the {\em Supersmoother} period-finding algorithm
\citep{fri84, rei94} on the source's $R$-band light curve. If Supersmoother
returned one or more periods in the 0.2-0.9 day range (typical of RR Lyrae
stars; \citealt{smi95}), then the variable source's light curve was phased
(period-folded) with each period and SDSS $r$-band RR Lyrae light curve
templates constructed by \citet{ses10a} were fitted to phased data.

The light-curve fitting was performed to minimize the robust goodness-of-fit
cost function defined in Equation~\ref{abs_sig_dev} in the least-square sense,
with the epoch of the peak brightness $\phi_0$, peak-to-peak amplitude $A_R$,
and peak brightness $R_0$ as free parameters. The quality of a template fit was
defined with a $\chi^2$-like parameter
\begin{equation}
\zeta = {\rm median}(|m_{observed}^i - m_{template}|/\epsilon_{observed}^i)\label{abs_sig_dev},
\end{equation}
where $m_{observed}$ and $\epsilon_{observed}$ are the observed $R$ band
magnitude and its uncertainty, $m_{template}$ is the magnitude predicted by the
template, and $i=1, N_{obs}$, where $N_{obs}$ is the number of observations.
Here we used the median to minimize the bias in $\zeta$ due to rare
observations with unreliable (non-Gaussian) errors (e.g., due to image
artifacts, cosmic rays). The template with the lowest $\zeta$ value was
selected as the best fit, and the best-fit parameters were stored. Finally, we 
visually inspected template-fitted light curves to tag sources as RR Lyrae
stars.

\begin{figure}
\epsscale{1.0}
\plotone{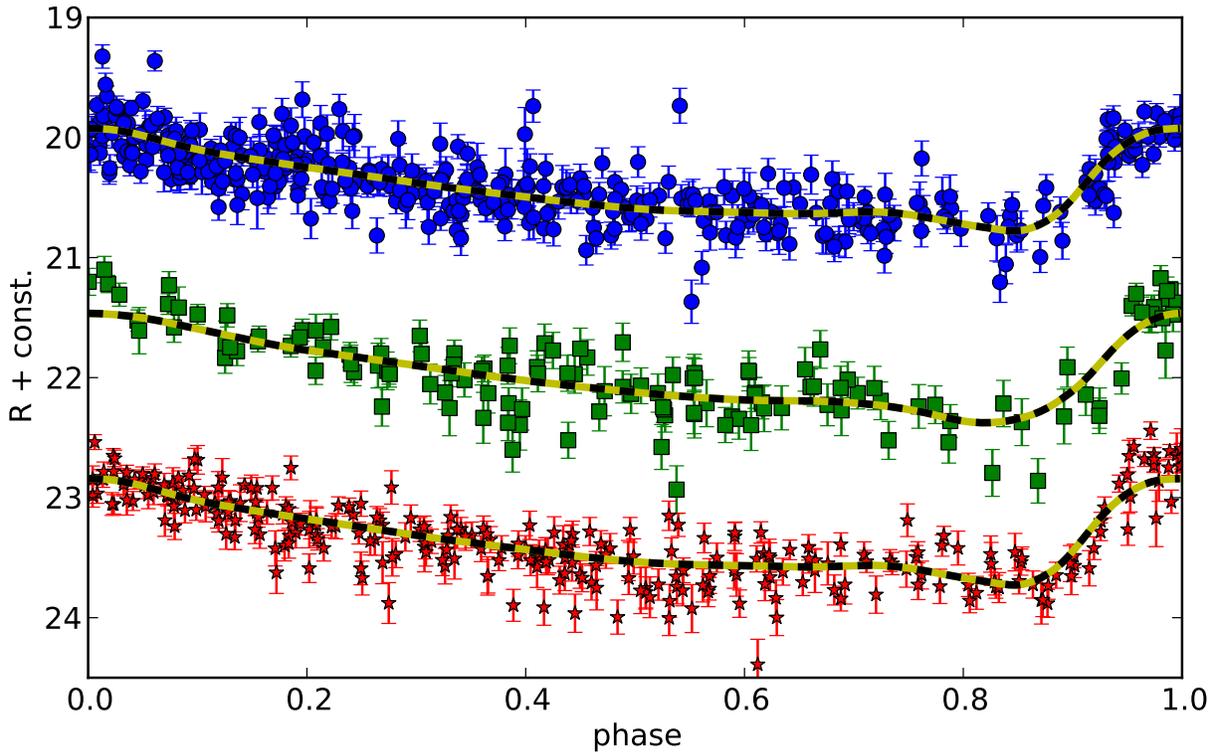}
\caption{
The phased $R$-band light curves of three RRab stars discussed in this work
(from top to bottom: RR5, RR8, and RR1). The light curves are offset for
clarity (offsets of 0, 1.5, and 3 mag from top to bottom) and the solid lines
show best-fit SDSS $r$-band templates constructed by \citet{ses10a}.
\label{fig2}} 
\end{figure}

The positions and light curve parameters of RR Lyrae stars observed in this
work are listed in Table~\ref{table-positions} and examples of phased $R$-band 
light curves are shown in Figure~\ref{fig2}. The periods and epochs of maximum
brightness are uncertain to within a few seconds and minutes, respectively,
allowing the phase of pulsation to be determined with better than 1\%
uncertainty. An accurate estimate of phase is important when subtracting the
velocity due to pulsations from the measured radial velocity, as we explain in
Section~\ref{radial_velocities}. In addition to 8 RR Lyrae stars discussed in
this work (RR1 to RR8), Table~\ref{table-positions} contains two RR Lyrae stars
that were spectroscopically followed-up after initial submission of this paper
(RR9 and RR10). These two stars were not used in the analysis presented below.
However, their position and velocities are mentioned in
Section~\ref{conclusions}, as these stars may indicate an eastward extension of 
the Cancer groups.

The heliocentric distances of RR Lyrae stars, $d$, are estimated from
measurements of their flux-averaged and extinction-corrected $R$-band
magnitudes, $\langle R \rangle$, and by assuming that $M_R \approx M_V$. The
$M_R \approx M_V$ approximation is uncertain at $\sim0.04$ mag level, as
estimated from the rms scatter of the difference of flux-averaged $r$ and $V$
band magnitudes of RR Lyrae stars from \citet{ses10a} (see their Table 3). The
absolute magnitudes of RRab stars in the Johnson $V$-band were calculated using
the \citet{chaboyer99} $M_V-[Fe/H]$ relation
\begin{equation}
M_V = (0.23\pm0.04)[Fe/H] + (0.93\pm0.12)\label{abs_mag},
\end{equation}
where ${\rm [Fe/H]}$ is the spectroscopic metallicity measured in
Section~\ref{metallicities}. The fractional error in the heliocentric distance
due to uncertainty in Equation~\ref{abs_mag}, spectroscopic metallicities,
and RR Lyrae evolution off the zero-age horizontal branch is estimated at 7\%
($\sim6$ kpc error in distance at 86 kpc). The apparent $\langle R \rangle$
magnitudes of the observed stars range from 19.9 mag to 20.3 mag, corresponding
to heliocentric distances of 77 kpc and 96 kpc, respectively.

\section{Completeness of the RR Lyrae Sample\label{completeness}}

The selection procedure described in the previous section yielded 8 RRab stars
at distances from 77 to 96 kpc from the Sun in the 45 deg$^2$ region of the sky
centered on the Praesepe cluster (see Figure~\ref{fig1}). The question we now
pose is whether these are all the RRab stars that could have been detected in
this distance range given the available PTF data. To answer this question we
need to know the success rate of our period-finding algorithm as the period
estimation is the most critical step in the selection of RR Lyrae stars. If the 
period of pulsation is not correctly estimated, the phased light curve will not
be identified as that of a RR Lyrae star.

To estimate the success rate of our period-finding algorithm, we have simulated
100 timeseries by randomly sampling 133 data points from a template RRab light
curve with a maximum brightness $R_0 = 20$ mag ($R_0$ of the faintest RRab star
in our sample), amplitude $A_R = 0.3$ mag, and period of 0.58 days. Random
sampling of a template light curve is a reasonable choice since the observed
light curves are almost uniformly sampled (see Figure~\ref{fig2}). The
amplitude of $A_R = 0.3$ mag represents the lower range of amplitudes that RRab
stars can have (e.g., see Figure 16 in \citealt{ses10a}) and the number of data
points corresponds to the number of observations of the least observed RRab
star in our sample (see Table~\ref{table-positions}). To make the light curves 
realistic, we added Gaussian noise to sampled magnitudes using a model that
describes the mean photometric uncertainty in the PTF $R$-band as a function of
magnitude. We then ran mock time-series through our period-finding algorithm
and counted time-series where the estimated period was within 1\% of the true
period.

Even for this worst-case scenario, we find that the periods are successfully
recovered in better than 95\% of cases and conclude that our selection
procedure has recovered almost every RRab star in the Praesepe region between
77 and 96 kpc from the Sun; at most one RRab star may have been missed.

\section{Spectroscopic Observations\label{spectro_obs}}

The group of 8 RR Lyrae stars detected in the Praesepe region at 86 kpc from
the Sun, and described in the previous section, is remarkable as it is the
densest and most numerous group that we have detected at these distances in our
preliminary search. However unlikely, this group could still be a chance
association of 8 distant stars and line-of-sight velocities are needed to test
whether the stars in this spatial group also form a moving group.

\subsection{Instrument Setup and the Observing Strategy}

The spectroscopic observations were obtained using two instruments: the Double
Spectrograph (DBSP; \citealt{og82}) mounted on the Palomar 5.1-m telescope and
the Low-Resolution Imager and Spectrograph (LRIS; \citealt{oke95}) mounted on
the Keck I telescope. A $1\arcsec$ wide slit, a 600 lines mm$^{-1}$ grating,
and a {5600 \AA} dichroic were used with both instruments. This instrument
setup covers a spectral range from 3800 {\AA} to {5700 \AA} and results in
resolutions of $R=1360$ and $R=1760$ for blue channels of DBSP and LRIS,
respectively.

The target RR Lyrae stars were observed over the course of several nights in
sets of one or two consecutive exposures followed by observations of a FeAr
(DBSP) or HgCdZn (LRIS) arc lamps. In order to avoid the discontinuity in the
radial velocity curve near the maximum light, the observations were scheduled to
target stars between phases of 0.1 and 0.85 of their pulsation cycle, with
earlier phases being preferred as the stars are brighter then.

In addition, on one night we also targeted a RRab star (AV Peg) previously
observed by \citet{lay94}. This star was observed in order to validate the
metallicity and systemic velocity measuring procedures described below. The log
of spectroscopic observations is provided in Table~\ref{table-data}.

\subsection{Data Reduction and Calibration\label{reduction}}

All data were reduced with standard IRAF\footnote{\url{http://iraf.noao.edu}}
routines, and spectra were extracted using an optimal (inverse
variance-weighted) method \citep{hor86}. The spectra of objects observed in two
exposures were reduced and calibrated independently and then combined. The
wavelength calibration of DBSP spectra was done using a single set of 49 FeAr
lines, while the LRIS spectra were calibrated using a set of 13 HgCdZn lines.
The highest root-mean-square (rms) scatter of the fit of the wavelength
calibration was {0.1 \AA} ($\sim6$ {\kms} at 4750 \AA) for DBSP spectra and
{0.05 \AA} ($\sim3$ {\kms} at 4750 \AA) for LRIS spectra.

To correct for the possible wavelength shift in DBSP spectra during an exposure
(e.g., due to instrument flexure), we measure the wavelengths
($\lambda_{\rm obs}$) of three ${\rm[Hg\, I]}$ sky lines and offset the
wavelength zero-point by the average of
$\Delta \lambda = \lambda_{\rm obs} - \lambda_{\rm lab}$, where
$\lambda_{\rm lab}=$ 4046.565, 4358.335, and {5460.750 \AA} are the laboratory
wavelengths of ${\rm [Hg\, I]}$ sky lines when observed in air. The uncertainty 
in the wavelength zero-point ($\sigma_{wave}$ in units of \AA) is estimated
using the standard deviation of $\Delta \lambda$.

To correct LRIS spectra we use the ${\rm [O\, I]}$ sky line at
$\lambda_{\rm lab}=5577.34$~{\AA}. As this line is outside the range of arc
lines used to calibrate LRIS spectra (the reddest arc line is at $\sim5460$
\AA), it is possible its position, and therefore the wavelength zero-point of
LRIS spectra, may be more uncertain if the {\em shape} of the wavelength
solution near 5577.34~{\AA} changes significantly throughout the night. This
would correspond to a distortion of the LRIS focal plane, not to flexure, which
is essentially a differential offset along this plane. To estimate the
uncertainty in the shape of the wavelength solution near 5577.34~{\AA}, we use
the following procedure. First, we evaluate on a fixed grid all wavelength
solutions obtained from arc images taken during a night. The solutions are
offset to the same wavelength zero-point by subtracting the average difference
between a solution and some reference solution. Then, we evaluate all solutions
at a single grid point for which the reference solution returns 5577.34~{\AA}.
We find the rms scatter around this point (5577.34~{\AA}) to be very similar to
the rms scatter of the fit of wavelength solutions ({$\sim0.05$ \AA} or 2-3
\kms). This means that the shape of the wavelength solution near 5577.34~{\AA}
does not vary significantly, even though this region is outside the range of arc
lines used in wavelength calibration. Therefore, for LRIS spectra we adopt the
rms scatter of the fit of the wavelength calibration as the uncertainty in the
wavelength zero-point ($\sigma_{wave}\sim0.05$ \AA). Later in
Section~\ref{radial_velocities}, we will use $\sigma_{wave}$ (in units of \AA)
to estimate the zero-point uncertainty in measured radial velocities
($\sigma_{zpt}$ in units of \kms).

\begin{figure}
\epsscale{0.9}
\plotone{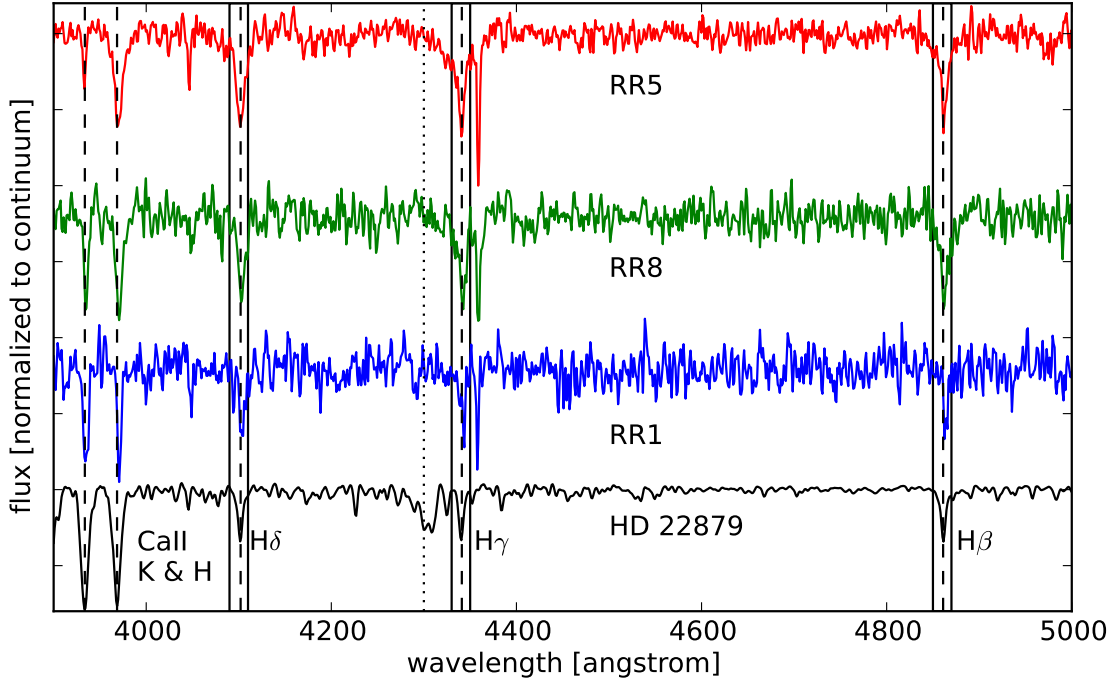}
\caption{
The DBSP spectra of some RRab stars discussed in this work, ordered from lowest
to highest S/N (top to bottom). An ELODIE template spectrum is shown at the
bottom of the panel. The DBSP spectra are in the observer's frame and the
ELODIE spectrum is in the rest frame. All spectra have been normalized to the
pseudo-continuum and have been offset in flux for clarity. The dashed lines
show the rest frame positions of, from left to right, ${\rm [Ca\, II]}$ K \& H
lines and Balmer lines (H$\delta$, H$\gamma$, H$\beta$ ). The boxes bracketing
Balmer lines show the spectral regions used for cross-correlation. The
absorption feature redward of the H$\gamma$ line is the {${\rm [Hg\, I]}$ 4358
\AA} sky-line that has been oversubtracted during extraction of spectra. The
dotted line shows the position of the CH G-band at 4300 \AA. Note how this
absorption feature is more pronounced in the spectrum of the ELODIE star than
in spectra of RR Lyrae stars. This indicates that the observed RR Lyrae stars
are likely more metal poor than the ELODIE star (a F9 dwarf star, with
${\rm [Fe/H]}=-0.84$ dex and $T_{eff} = 5841$ K).
\label{fig3}} 
\end{figure}

The signal-to-noise ratio (S/N) of the final spectra at {4750 \AA} ranged from
10 to 20 for DBSP spectra and 20 to 30 for LRIS spectra. Figure~\ref{fig3}
shows a few examples of observed DBSP spectra, ordered from the highest to
lowest S/N. The Balmer lines (starting at H$\beta$) and ${\rm [Ca\, II]}$ H and
K lines are easily seen in the spectra. The CH G-band absorption feature at
{4300 \AA} is relatively weak in spectra of RR Lyrae stars, indicating that the 
RR Lyrae stars shown in Figure~\ref{fig3} are likely more metal poor than the
ELODIE template star HD 22879 (${\rm [Fe/H]}= -0.84$ dex). The ELODIE template
star HD 22879 is a F9 dwarf ($T_{eff} = 5841$ K) and is one of the radial
velocity standard stars used in the next section.

\subsection{Line-of-Sight Velocities\label{radial_velocities}}

Heliocentric radial velocities, $v_r$, were obtained by cross-correlating our
spectra with spectra of 11 F2 to F9 dwarf stars (hereafter template spectra,
see Table~\ref{elodie_templates}) taken from the ELODIE
catalog\footnote{\url{http://atlas.obs-hp.fr/elodie/}} \citep{moultaka04}. The
template spectra were degraded to the resolution of our observations and were
selected to encompass the range of spectral types covered by RRab stars as they
pulsate. Expanding the list of template spectra to include more stars and more
spectral types did not significantly reduce the uncertainty of measured radial
velocities.

The cross-correlation was performed using the FXCOR task in IRAF, with template
and science spectra normalized to the pseudo-continuum prior to
cross-correlation. For our relatively low resolution spectra, the narrowest
cross-correlation peaks were obtained by restricting the correlation region
in template and science spectra to cores and wings of Balmer lines (spectral
regions used in cross-correlation are shown in Figure~\ref{fig3}). Addition of
other spectral regions or Fourier filtering did not significantly improve the
cross-correlation.

The radial velocities were measured by fitting a Gaussian to the top 25\% of
the cross-correlation peak and the final $v_r$ radial velocity of an object was
calculated as the average of the 11 $v_r$ values obtained from
11 cross-correlations with ELODIE templates, where each $v_r$ value was
weighted by its cross-correlation error $\sigma_{cc}$. The cross-correlation
error $\sigma_{cc}$ reported by FXCOR includes the uncertainties due to
broadening of spectral features during an exposure, random errors in wavelength
calibration, and errors due to ELODIE template mismatch. Therefore, radial
velocities from better template fits contribute more towards the final radial
velocity value.

In addition to the cross-correlation error, there is an error in radial velocity
due to the uncertainty in determining the wavelength zero-point. We calculate
this zero-point error as
\begin{equation}
\sigma_{zpt}=\frac{\sigma_{wave}c}{4750}
\end{equation}
in units of km $s^{-1}$, where $\sigma_{wave}$ is the uncertainty in wavelength
zero-point in units of Angstr\"om (see Section~\ref{reduction}). The final $v_r$
radial velocities and their uncertainties (cross-correlation and zero-point) are
listed in Table~\ref{table-data}.

\begin{figure}
\epsscale{0.5}
\plotone{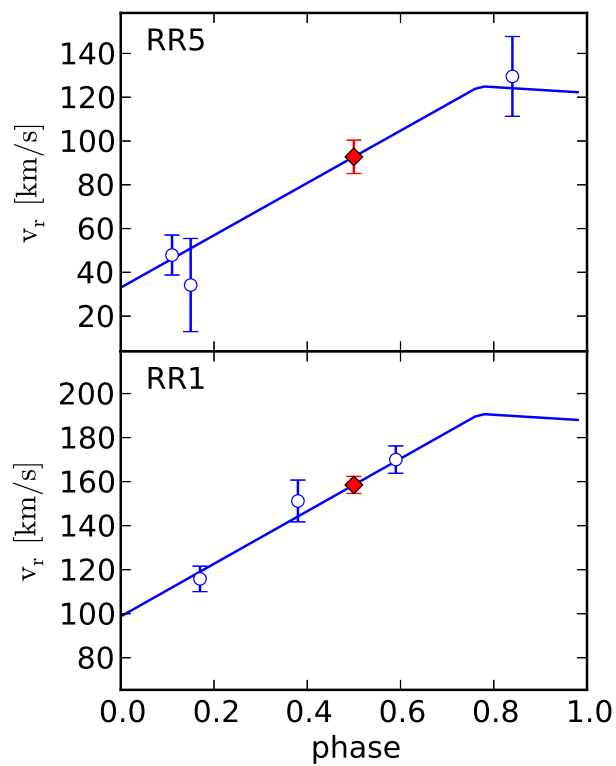}
\caption{
Fits of the radial velocity template of X Ari ({\em solid line}) to two stars
in our sample. The open circles show individual $v_r$ measurements and their
error bars indicate the uncertainty (cross-correlation and zero-point errors
added in quadrature). The red diamond symbol at phase 0.5 shows the systemic
velocity calculated as the weighted average of $v_\gamma$ values, where the
$v_\gamma$ values were obtained by fitting the model template to each of the
observed radial velocities. The uncertainty in the systemic velocity includes
the observational errors (cross-correlation and zero-point), and model errors.
\label{fig4}} 
\end{figure}

Following \citet{hb85}, the systemic (center-of-mass) velocity of RR Lyrae
stars, $v_\gamma$, was obtained by fitting the radial velocity curve of the
RRab star X Ari to observed heliocentric velocities. We used Layden's (1994)
parametrization (see his Figure 8) of the radial velocity curve that
\citet{oke66} measured from the H$\gamma$ line. This template is an appropriate
choice for fitting radial velocities measured using Balmer lines (i.e., as done
in this work). Radial velocity templates based on measurements of metallic
lines, such as the one presented in \citet{liu91}, would not be appropriate
here as radial velocity curves of metallic and Balmer lines have different
shapes and amplitudes \citep{oke66}. The X Ari template was shifted in velocity
to match the measurement for each star at the corresponding phase, and the
systemic velocity is that of the shifted template at phase 0.5. Two examples of
template fits are shown in Figure~\ref{fig4}.

Finally, the error calculation for the systemic velocity includes terms that
take into account errors in the model radial velocity curve. Following
\citet{vzg05} (see their Section 3), we use the expresion given below in which
the third term accounts for likely uncertainties in the phase where the
velocity curve passes through the systemic velocity, and the fourth term is
related to possible variations in the slope of the velocity curve:
\begin{equation}
\sigma_{\gamma}^2 = \sigma_{cc}^2 + \sigma_{zpt}^2 + (119.5\times0.1)^2 + (23.9\Delta\phi)^2
\label{eq-sigma}
\end{equation}
where $\Delta\phi = \phi_{obs} - 0.5$, and $\phi_{obs}$ is the phase of
observation. The $\sigma_{cc}$ and $\sigma_{zpt}$ are the cross-correlation and
zero-point error, respectively. The contributions of model uncertainties are of
the order of $\sim15$ {\kms} and are higher than the observed radial velocity
errors ($\sigma_{cc}$ and $\sigma_{zpt}$ added in quadrature). Therefore, to
reduce the uncertainty in the systemic velocity we need to reduce the
uncertainty introduced by the model, that is, we should observe the radial
velocity curve more than once.

When there is more than one $v_\gamma$ estimate available for a RRab star due to
multiple measurements, we simply calculate the weighted average of $v_\gamma$,
$\langle v_\gamma\rangle$, and its associated uncertainty. Finally, we correct
the heliocentric systemic velocities $\langle v_\gamma\rangle$ for the solar
motion with respect to the Galactic center using
\begin{equation}
\langle v_{gsr} \rangle = \langle v_\gamma\rangle +v_U\cos b \cos l + (v_V+v_{LSR})\cos b \sin l + v_W \sin b,
\end{equation}
where $(v_U, v_V, v_W) = (10.0, 5.2, 7.2)$ {\kms } and $v_{LSR}=220$ {\kms}
\citep{bm98}, where $v_{LSR}$ is the velocity of the local standard of rest.
These velocities in the Galactic rest frame ($\langle v_{gsr} \rangle$) are
adopted in the rest of this paper.

To validate the above procedure, we measured the radial velocity of AV Peg. The
light curve parameters for this star were taken from the GEOS RR Lyrae database
\citep{leb07}. After correcting the observed heliocentric velocity for
pulsation using the model radial velocity curve, we obtained
$\langle v_\gamma\rangle = -64.8\pm18.7$ {\kms}. This value agrees within the
uncertainties with the $\langle v_\gamma\rangle = -58\pm1$ {\kms} estimated by
\citet{lay94} (see his Table 9).

\subsection{Metallicities\label{metallicities}}

Spectroscopic metallicities were measured following the method and calibration
of \citet{lay94} which involves comparing the pseudo-equivalent width of
${\rm [Ca\, II]}$ K line, W(K), against the mean pseudo-equivalent widths of
$\beta$, $\gamma$, and $\delta$ Balmer lines, W(H). 

The pseudo-equivalent widths of the ${\rm [Ca\, II]}$ K line and Balmer lines
were measured using the EWIMH program\footnote{\url{http://physics.bgsu.edu/~layden/ASTRO/DATA/EXPORT/EWIMH/ewimh.htm}} (written by A.~Layden) from DBSP and
LRIS spectra normalized to the pseudo-continuum. Measured pseudo-equivalent
widths ({\rm $W^\prime(K)$}, {\rm $W^\prime(H\delta)$},
{\rm $W^\prime(H\gamma)$}, and {\rm $W^\prime(H\beta)$}) were transformed to the
Layden's (1994) pseudo-equivalent width system using
\begin{eqnarray}
W(K) = 1.23 W^\prime(K) - 0.44\label{K_DBSP}\\
W(H\delta) = 0.84 W^\prime(H\delta) + 1.64 \\
W(H\gamma) = 1.13 W^\prime(H\gamma) - 0.63 \\
W(H\beta) = 1.02 W^\prime(H\beta) + 0.98
\end{eqnarray}
for pseudo-equivalent widths measured from DBSP spectra, and using
\begin{eqnarray}
W(K) = 1.16 W^\prime(K) - 0.39 \\
W(H\delta) = 1.02 W^\prime(H\delta) + 1.39 \\
W(H\gamma) = 1.07 W^\prime(H\gamma) - 0.45 \\
W(H\beta) = 1.01 W^\prime(H\beta) + 1.18\label{Hb_LRIS}
\end{eqnarray}
for pseudo-equivalent widths measured from LRIS spectra. To derive these
equations we observed 5 of the equivalent width standard stars of \citet{lay94}
with DBSP and four more with LRIS; see Table~\ref{EW_stds}. Comparing our
measurements of their pseudo-equivalent widths with those listed in
\citet{lay94} yielded Equations~\ref{K_DBSP} to~\ref{Hb_LRIS}. After the
transformation, the W(K) was corrected for interstellar ${\rm [Ca\, II]}$
absorption using the \citet{bee90} model. This correction decreases the
measured metallicities by $\sim0.15$ dex.

The spectroscopic metallicity was calculated as
\begin{equation}
[Fe/H] = \frac{W(K) - a - bW(H)}{c+dW(H)}\label{spec_feh},
\end{equation}
where $a = 13.858$, $b = -1.185$, $c = 4.228$, $d = -0.32$ (see Table 8 in
\citealt{lay94}). Equation~\ref{spec_feh} was obtained by inverting Equation 7
from \citet{lay94}. The metallicities obtained using this method are listed in
Table~\ref{table-data}.

For the most part, ${\rm [Fe/H]}$ estimates obtained from multiple observations
of the same star are consistent within 0.1 to 0.2 dex, even for RR1 which was
observed with different instruments (DBSP and LRIS). This scatter is consistent
with the systematic uncertainty of this method, which is $\sim0.15$ dex
\citep{lay94}. The exception to this is star RR3 for which we measured
${\rm [Fe/H]} = -1.5$ dex from a DBSP spectrum and ${\rm [Fe/H]} = -2.0$ dex
from a LRIS spectrum. If we assume ${\rm [Fe/H]} = -1.8$ dex as the metallicity
of this star (the average of -1.5 and -2.0), we find that the two measurements
are still within $2\sigma$ of the average value, which is statistically not
improbable. While it may be tempting to assign lower significance to DBSP
metallicity measurements due to lower S/N of DBSP spectra (10 to 20 vs.~$>20$
for LRIS spectra), multiple metallicity measurements of RR5 argue against that.
RR5 has been observed with DBSP over three nights, its spectra have S/N in the
10 to 20 range, and yet the standard deviation of its metallicity measurements
is 0.1 dex, consistent with the systematic uncertainty.

To validate the zeropoint of our metallicity scale, we measured ${\rm [Fe/H]}$
of RRab star AV Peg observed by \citet{lay94}. We obtained ${\rm [Fe/H]}=0.1$
dex while \citet{lay94} cites ${\rm [Fe/H]}=-0.1$ dex. These two values are
roughly within the systematic uncertainty of this method, indicating that our
metallicities are on the same zeropoint system as the \citet{lay94} stars
(i.e., the \citealt{zw84} globular cluster abundance scale; \citealt{lay94}).

\section{Discussion\label{discussion}}

In this section we address the following questions.
\begin{enumerate}
\item What is the mean velocity and velocity dispersion of our sample, and how
do these values compare to the literature?
\item Is the observed distribution of velocities a Gaussian or a non-Gaussian
distribution?
\item Are the observed velocity groups bound or unbound systems?
\item What does the metallicity of their stars say about the progenitor(s)?
\item Are the observed velocity groups related to known halo substructures?
\end{enumerate}

The galactocentric rest-frame velocities $\langle v_{gsr} \rangle$ listed in
Table~\ref{table-data} indicate that the observed RR Lyrae stars can be split
into two velocity groups, one moving at $\bar{v}_{gsr} = 78.0\pm5.6$ {\kms} and
the other one at $\bar{v}_{gsr} = 16.3\pm7.1$ {\kms}, where $\bar{v}_{gsr}$ is
the weighted mean of $\langle v_{gsr} \rangle$ values listed in
Table~\ref{table-data}. The uncertainty in $\bar{v}_{gsr}$ is calculated as the
standard error of the weighted mean. We tentatively name these moving groups
Cancer group A and B. The weighted mean velocity for the entire sample is
$\bar{v}_{gsr} = 53.5\pm4.4$ {\kms}.

To estimate the velocity dispersion of these moving groups, we use the
following Monte Carlo procedure. We generate 1000 sets where each set contains
four mock line-of-sight velocities, $v_{mock}$ (four values because there are
four RRab stars in each moving group). A $v_{mock}$ value is generated by
drawing a single value from a Gaussian distribution centered on the
$\langle v_{gsr} \rangle$ value of a star (e.g., 75.6 {\kms} for RR1) and
having a width equal to the uncertainty on $\langle v_{gsr} \rangle$ (e.g., 8.5
{\kms} for RR1). The weighted standard deviation is calculated and stored
for each set, with $v_{mock}$ values weighted by uncertainty on corresponding
$\langle v_{gsr} \rangle$ values listed in Table~\ref{table-data}. The velocity
dispersion is then the average value of the distribution of 1000 weighted
standard deviations and the uncertainty of velocity dispersion is the standard
deviation of this distribution. Using this procedure we find that the two
moving groups have velocity dispersions of $\sigma_{v_{gsr}}=12.4\pm5.0$ {\kms}
(group A) and $\sigma_{v_{gsr}}=14.9\pm6.2$ {\kms} (group B). The velocity
dispersion of the entire sample is $\sigma_{v_{gsr}}= 37.7\pm5.0$ {\kms}.

The measured velocity dispersion for the full sample is inconsistent with the
halo velocity dispersion profile obtained by \citet{brown10} using a sample of
910 distant halo stars. At 92 kpc, the extrapolation of the \citet{brown10}
model (see their Equation 6) would describe the velocity distribution of stars
as a 68 {\kms}-wide Gaussian centered on 0 {\kms}. The mean velocity of our full
sample is at least $10\sigma$ away from 0 {\kms}, and the velocity dispersion of
our full sample is at least $5\sigma$ lower than the velocity dispersion
extrapolated from the \citet{brown10} model. One possible explanation of this
discrepancy is that \citet{brown10} Equation 6 cannot be extrapolated outside
the $16 < R_{GC} < 64$ kpc range within which the relation is assumed to be
valid \citep{brown10}. A more likely explanation is that we are observing two
moving groups with their own internal kinematics (mean velocity and velocity
dispersion). If that is the case, a comparison of the predicted and measured
velocity dispersion for the full sample is not a sensible thing to do.

The question of whether the distribution of velocities listed in
Table~\ref{table-data} is a Gaussian or a non-Gaussian distribution is an
important one, as deviations from normality are usually interpreted as a
signature of velocity groups in the smooth halo component
\citep{harding01,duffau06,vivas08}. To test for the presence of velocity groups
in the halo, \citet{harding01} recommend the \citet[][SW]{shapiro65}
statistical test of normality to be applied to velocity histograms. This test
is sensitive to many different deviations from the Gaussian shape and does not 
depend on the choice of mean or dispersion of the normal distribution (see
\citealt{harding01} Section 5.1 and \citealt{ds86} for more details on the
test).

\begin{figure}
\epsscale{0.5}
\plotone{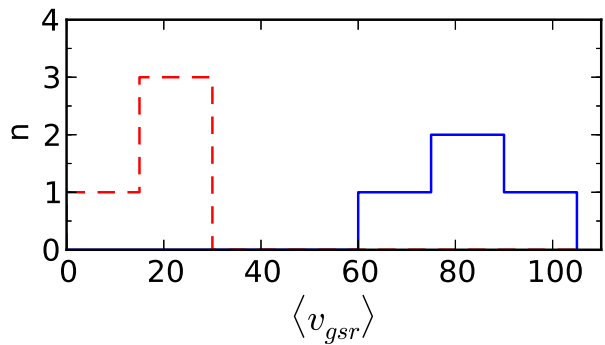}
\caption{
Histogram of $\langle v_{gsr}\rangle$ of RR Lyrae stars in Cancer groups. The
bin size is 15 km s$^{-1}$, which is about the average velocity error in
$\langle v_{gsr}\rangle$. The distribution of observed velocities seems to be
bimodal, with two velocity peaks centered on $\langle v_{gsr}\rangle=78$ {\kms}
and $\langle v_{gsr}\rangle=16$ {\kms}, respectively. See
Section~\ref{discussion} for a discussion of statistical significance of this
bimodality.
\label{fig5}} 
\end{figure}

Following the procedure detailed in \citet[][see their Section 3]{ses10b} that
takes into account uncertainties in velocities, we use the SW test and find
that there is a 37\% probability that the observed sample of velocities was
drawn from a single Gaussian distribution. However, the SW test does not have
the ability to assess the rather peculiar nature of the {\em shape} of the
observed distribution (i.e., it cannot distinguish the particular nature of
non-Gaussianity). To include this information, we consider the 37\% chance that
the measured velocities were drawn from a Gaussian distribution, and
estimate the probability that this chance would produce the observed
distribution seen in Figure~\ref{fig5}. We generate 100,000 samples and for
each sample draw 8 velocities from a Gaussian centered on 53.5 km s$^{-1}$ and
37.7 km s$^{-1}$ wide (i.e., the velocity distribution of the full sample of RR
Lyrae stars observed in the Cancer region). Each sample is then divided into
stars with velocities greater and smaller than 53.5 km s$^{-1}$ (``positive''
and ``negative'' groups), and the mean velocities and velocity dispersions are
calculated for each group ($v_{pos}$, $v_{neg}$, $\sigma_{pos}$, and
$\sigma_{neg}$). In only $\sim1.6\%$ of generated samples, we find that the
``positive'' and ``negative'' groups have mean velocities and velocity
dispersions in ranges covered by Cancer groups ($72 < v_{pos} < 84$ and
$7 < \sigma_{pos} < 18$, and $9 < v_{neg} < 24$ and $8 <\sigma_{neg}<22$). The
percentage of generated samples with mean velocities and velocity dispersions in
ranges covered by Cancer groups is even lower ($0.02\%$), if the samples are
drawn from a 68 km s$^{-1}$-wide Gaussian centered on 0 km s$^{-1}$ (i.e., the
velocity distribution at 92 kpc as extrapolated from Equation 6 by
\citealt{brown10}). Therefore, even if the observed velocities were drawn from a
Gaussian distribution (a 37\% chance), it is unlikely they would create the
distribution seen in Figure~\ref{fig5}, as the chance of that happening is only 
about 0.6\% ($0.37\times0.016 \sim 0.006$), if the velocity distribution at 92
kpc is a 37.7 km s$^{-1}$-wide Gaussian centered on 53.5 km s$^{-1}$, or about
$0.07\%$ ($0.37\times0.0002\sim 7\cdot10^{-5}$), if the velocity distribution
follows the extrapolation of the \citet{brown10} model. The next simplest
explanation for the observed distribution of velocities is that we are observing
two velocity groups, moving at $\bar{v}_{gsr} = 78.1\pm5.7$ {\kms} and
$\bar{v}_{gsr} = 16.3\pm7.1$ {\kms} and having velocity dispersions of
$\sigma_{v_{gsr}}=12.4\pm5.0$ {\kms} and $\sigma_{v_{gsr}}=15.0\pm6.1$ {\kms}.

These estimated velocity dispersions are consistent with velocity dispersions
of dwarf spheroidal galaxies \citep[][e.g., Leo II]{wal07} and globular
clusters \citep[][2010 edition, e.g., NGC 362]{har96}. Due to significant
uncertainities in the velocities, these velocity dispersions are most likely
upper limits on the intrinsic velocity dispersions of the Cancer groups. While
the velocity dispersions of the two groups are consistent with velocity
dispersions of dwarf spheroidal galaxies and globular clusters, the groups are
spatially quite extended (the maximum separation between members is several kpc)
making it very unlikely that they are bound systems, and are more likely to be
debris of tidally disrupted dwarf galaxies or globular clusters.

The metallicity of these groups may provide more information about their
progenitors. The first group has a median metallicity of ${\rm [Fe/H] = -1.6}$
dex and the second group has ${\rm [Fe/H] = -2.1}$ dex on the \citet{zw84}
globular cluster abundance scale, indicating that the progenitors of these two
substructures are systems with old, metal-poor populations. The metallicity
spread (standard deviation) in both groups is rather large ($\sim0.4$ dex) and
a few times greater than the uncertainty in individual metallicity estimates
($\sim0.15$ dex). Such a large dispersion in ${\rm [Fe/H]}$ indicates
self-enrichment and implies the progenitor structures were dwarf galaxies and
most likely not globular clusters. Using the metallicity-luminosity relation by
\citet{kir11} (see their Equation 9), we can place a lower limit on the
luminosity of group A and B progenitors. Taking into account the intrinsic
spread in this relation, we obtain $L\gtrsim5.5\times10^5$ L$_\sun$ for the
group A progenitor and $L\gtrsim9.1\times10^3$ L$_\sun$ for the group B
progenitor. The luminosity limit of $L\gtrsim5.5\times10^5$ L$_\sun$ indicates
that the group A progenitor was likely more luminous than one of the classical
dwarf spheroidal galaxies, such as Leo II (e.g., see Figure 3 by
\citealt{kir11}). 

However, we do note that there are only four stars in each group and that the
metallicity spread can change drastically if a single measurement is not
considered when estimating the spread. For example, if the metallicity of RR8 is
not taken into account when estimating the metallicity spread of the second
group, the new spread would be $\sim0.1$ dex which would be consistent with
measurement uncertainties and consistent with a single stellar population
(i.e., a globular cluster as the progenitor). Additional observations of RR
Lyrae stars on the metal-poor and metal-rich ends (e.g., RR2, RR3, and RR8)
would be very useful as they would further test the validity of the estimated
metallicity dispersions.

\citet{sha10} have found two overdensities of M giants extending into this
region of the sky, one at $22.6\pm11$ kpc (their group A13) and the other one
at $96.6\pm48$ kpc (their group A9). Both of these groups have density peaks
(centers) that are offset by at least $20\arcdeg$ from the location of Cancer
groups discussed here, with the group A13 peaking at ${\rm l}\sim144\arcdeg$
and ${\rm b}\sim31\arcdeg$ and group A9 peaking at ${\rm l}\sim188\arcdeg$ and
${\rm b}\sim20\arcdeg$. The Cancer groups are likely not associated with the
\citet{sha10} group A13 based on estimated distances of M giants provided by
\citet{sha10} ($22.6\pm11$ kpc, see their Table 2). Based on the distance alone 
($96.6\pm48$ kpc), it is not impossible that the Cancer groups may be associated
with the \citet{sha10} group A9. However, we do note that the Cancer groups
discussed in this work are on the very edge of the \citet{sha10} group A9 (see
their Figure 7) and that the distance of group A9 has uncertainty of $\sim50$
kpc.

\citet{new03} have found an overdensity of A-type stars about $10\arcdeg$ north
in declination from Cancer groups and at a similar distance ($\sim90$ kpc from
the Sun; see their Figure 1). At a similar location, \citet{ruh11} have found
``a very faint indication of an overdensity'' of blue horizontal branch (BHB)
stars. Both studies speculated that the overdensities they observed were
associated with the Sagittarius tidal debris. However, recent modelling of the
Sagittarius streams by \citet{lm10} suggests that there should not be any debris
at these distances. It is unclear whether these overdensities and Cancer groups
are related due to lack of spectroscopic follow-up of BHB and A-type stars in
\citet{new03} and \citet{ruh11} overdensities.

The surface density of RRab stars in Cancer groups A and B seems to be similar
to that of the Pisces Overdensity/Stream ($\sim0.3$ RRab per deg$^2$), a halo
substructure located at a similar distance ($\sim85$ kpc) from the Galactic
Center \citep{ses07,wat09,ses10a}. In addition, both of these substructures
seem to have two velocity groups (the Pisces Overdensity has groups moving at
50 {\kms} and $-52$ {\kms}; \citealt{kol09, ses10b}). The morphology of both
substructures is reminiscent of debris clouds observed in some simulations
(e.g., \citealt{bj05, joh08}). According to a recent study by \citet{joh12},
the clouds represent debris that is slowly turning around at apocenters of
orbits that are typically more eccentric than orbits of debris streams, such as
the Sagittarius stream. If so, then ``there must be low-density stellar streams
moving between these {\em apocentric clouds} and passing through the inner
Galaxy at high speed.'' \citep{joh12}. By following up RR Lyrae stars selected
from PTF and other surveys, it will be possible to identify such low density
streams, if they exist, and to explore their relationship with Cancer groups A
and B.

\section{Conclusions\label{conclusions}}

We confirm the existence of two kinematic groups in the direction of the Cancer
constellation (${\rm R.A}\sim129\arcdeg$ and ${\rm Dec}\sim20\arcdeg$, or
${\rm l}\sim205\arcdeg$ and ${\rm b}\sim32\arcdeg$), located at 92 kpc from the
Galactic center (86 kpc from the Sun). These groups, tentatively named Cancer
groups A and B, are moving at $\bar{v}^A_{gsr} = 78.0\pm5.7$ {\kms} (Cancer
group A) and $\bar{v}^B_{gsr} = 16.3\pm3.8$ {\kms} (Cancer group B). The groups
have velocity dispersions smaller than 15 {\kms}, are spatially extended (about
several kpc), metal-poor (median metallicities of ${\rm [Fe/H]^A = -1.6}$ dex
and ${\rm [Fe/H]^B=-2.1}$ dex), and have a metallicity spread of $\sim0.4$ dex.
These results suggest that the observed groups are debris of tidally disrupted
dwarf galaxies, possibly near the apocenters of their orbits. Whether these
groups are related to known substructures in the Galactic halo is unclear at
this point, and answering this question may require extensive orbit modeling and
comparisons with simulations (e.g., as done by \citealt{joh12} and
\citealt{car12}).

Observations of two RR Lyrae stars (RR9 and RR10 in Tables~\ref{table-positions}
and~\ref{table-data}) obtained after this paper was submitted for peer-review,
may help with the modeling of Cancer groups' orbits. The two RR Lyrae stars have
velocities of 88 {\kms} and 38 {\kms} that are consistent (to within the
uncertainties) with the mean velocities of Cancer groups A and B, and are at
similar distances ($\sim76$ kpc from the Sun). They are offset $\sim25\arcdeg$
east of the Cancer groups and may indicate an eastward extension of these
groups. However, due to the sparse coverage in PTF of the sky between Cancer
groups and RR Lyrae stars RR9 and RR10, we are unable to verify at this point
whether RR9 and RR10 are truly related to the Cancer groups or not.

Initially, this work was motivated by a hypothesis that distant halo
substructures may be found by simply looking for distant spatial groups of RR
Lyrae stars. Based on this work and previous work by \citet{kol09} and
\citet{ses10b}, we conclude that this is indeed an efficient approach to
finding and following-up halo substructures. So far, all of the distant
(galactocentric distances greater than 80 kpc) RR Lyrae stars located in spatial
groups have proven to be members of a halo substructure (e.g., Cancer groups in
this work and RR Lyrae stars in the Pisces Overdensity;
\citealt{kol09, ses10b}). Since the density profile of the relatively smooth,
inner halo steepens beyond 30 kpc \citep{sji11, dbe11}, it may be that the
majority, if not all, of RR Lyrae stars beyond $\sim30$-$40$ kpc are part of
some halo substructure. Spectroscopic followup of distant RR Lyrae stars not
studied in this work may provide more data to support or refute this hypothesis,
and we plan to follow this strategy with other RR Lyrae stars observed by the
Palomar Transient Factory.

\acknowledgments

J.G.C. and B.S. thank NSF grant AST-0908139 to J.G.C for partial support, as do
S.R.K (to NSF grant AST-1009987), and C.J.G (for a NASA grant). Support for
this work was provided by NASA through Hubble Fellowship grant 51256.01 awarded
to E.N.K by the Space Telescope Science Institute, which is operated by the
Association of Universities for Research in Astronomy, Inc., for NASA, under
contract NAS 5-26555.

We thank the referee for a thorough review and suggestions which led to an
improved manuscript. B.S. would like to thank \v{Z}.~Ivezi\'c, B.~Willman, and
K.~Vivas for useful discussions. We thank I.~Arcavi, A.~Gal-Yam, P.~Groot,
A.~Horesh, and D.~Perley for observing at Keck and Palomar. We thank the staff
at the Palomar Hale telescope for help and support with observations. We are
grateful to the many people who have worked to make the Keck Telescope and its
instruments a reality and to operate and maintain the Keck Observatory. The
authors wish to extend special thanks to those of Hawaiian ancestry on whose
sacred mountain we are privileged to be guests. Without their generous
hospitality, none of the observations presented herein would have been
possible.

This article is based on observations obtained with the Samuel Oschin Telescope
as part of the Palomar Transient Factory project, a scientific collaboration
between the California Institute of Technology, Columbia University, Las Cumbres
Observatory, the Lawrence Berkeley National Laboratory, the National Energy
Research Scientific Computing Center, the University of Oxford, and the Weizmann
Institute of Science.

\begin{deluxetable}{llrrcrrrrr}
\rotate
\tabletypesize{\scriptsize}
\tablecolumns{10}
\tablewidth{0pc}
\tablecaption{Positions and Light Curve Parameters of RR Lyrae Targets\label{table-positions}}
\tablehead{
\colhead{ID$^a$} & \colhead{IAU name} & \colhead{R.A.$^b$} & \colhead{Dec$^b$} &
\colhead{distance$^c$} & \colhead{$A_R^d$} &
\colhead{$R_0^e$} & \colhead{Period$^f$} &
\colhead{HJD$_0^g$} & \colhead{$N_{obs}^h$} \\
\colhead{} & \colhead{} & \colhead{(deg)} & \colhead{(deg)} & \colhead{(kpc)} &
\colhead{(mag)} & \colhead{(mag)} & \colhead{(d)} &
\colhead{(d)} & \colhead{}
}
\startdata
RR1  & PTF1J084844.61+202430.8 & 132.185880 & 20.408611 & 87.1/93.2  & 0.887 & 19.84 & 0.515257 & 55303.686299 & 291 \\
RR2  & PTF1J084313.05+193901.1 & 130.804370 & 19.650306 & 78.2/84.4  & 0.889 & 19.81 & 0.556328 & 55271.737376 & 414 \\
RR3  & PTF1J083851.75+221641.5 & 129.715656 & 22.278200 & 85.9/92.2  & 0.766 & 19.82 & 0.555729 & 55590.910809 & 160 \\
RR4  & PTF1J082914.96+185607.4 & 127.312372 & 18.935392 & 86.0/92.4  & 0.629 & 20.01 & 0.594511 & 55667.607592 & 193 \\
RR5  & PTF1J083636.44+200954.9 & 129.151846 & 20.165259 & 96.0/102.3 & 0.855 & 19.92 & 0.610057 & 55594.666378 & 406 \\
RR6  & PTF1J083339.04+192641.5 & 128.412679 & 19.444856 & 92.0/98.4  & 0.697 & 19.97 & 0.633480 & 55823.965992 & 271 \\
RR7  & PTF1J083033.30+220217.9 & 127.638788 & 22.038294 & 77.0/83.5  & 0.703 & 19.64 & 0.545361 & 55594.799336 & 400 \\
RR8  & PTF1J082333.11+213414.6 & 125.887964 & 21.570710 & 84.2/90.8  & 0.912 & 19.96 & 0.517656 & 55542.999049 & 133 \\
\hline
RR9  & PTF1J103051.69+202330.8 & 157.715375 & 20.391878 & 75.5/79.3  & 0.914 & 19.35 & 0.544278 & 55515.539757 & 82  \\
RR10 & PTF1J102040.66+212315.1 & 155.169417 & 21.387514 & 76.1/80.2  & 0.759 & 19.52 & 0.611088 & 55244.324620 & 96  \\
\hline
AV Peg &         AV Peg        & 328.011644 & 22.574827 &  0.7/7.9   & 1.040 &  n.a. & 0.390381 & 55833.513400 & n.a.
\enddata
\tablecomments{The first 8 RR Lyrae stars are located in the Cancer region, the
next two are offset by $\sim25\arcdeg$ east from the Cancer region and suggest
an eastward extension of Cancer groups, and AV Peg is a known RR Lyrae stars
used for validation of various procedures used in this work.}
\tablenotetext{a}{Short name.}
\tablenotetext{b}{Equatorial J2000.0 right ascension and declination from SDSS DR8 catalog.}
\tablenotetext{c}{Heliocentric/galactocentric distance calculated using spectroscopic metallicity.}
\tablenotetext{d}{{Light} curve amplitude in the Mould-$R$ band ($V$-band amplitude for AV Peg).}
\tablenotetext{e}{$R$-band magnitude at the epoch of maximum brightness.}
\tablenotetext{f}{Period of pulsation.}
\tablenotetext{g}{Reduced Heliocentric Julian Date of maximum brightness (HJD - 2400000).}
\tablenotetext{h}{Number of observations in the Mould-$R$ band.}
\end{deluxetable}

\clearpage

\begin{deluxetable}{lrcrrrrrrr}
\rotate
\tabletypesize{\scriptsize}
\tablecolumns{10}
\tablewidth{0pc}
\tablecaption{Observing Log, Velocities, and Metallicities\label{table-data}}
\tablehead{
\colhead{ID$^a$} & \colhead{HJD$_{\rm spectrum}^b$} &
\colhead{Exposures$^c$} & \colhead{Phase$^d$} &
\colhead{$v_r\pm\sigma_{cc}\pm\sigma_{zpt}^e$} &
\colhead{$v_\gamma\pm\sigma_{\gamma}^f$} &
\colhead{$\langle v_\gamma\rangle^g$} &
\colhead{$\langle v_{gsr}\rangle^h$} &
\colhead{${\rm [Fe/H]^i}$} \\
\colhead{} & \colhead{(d)} &
\colhead{(s)} & \colhead{} & \colhead{(\kms)} &
\colhead{(\kms)} & \colhead{(\kms)} & \colhead{(\kms)}
& \colhead{(dex)} 
}
\startdata
RR1  & 55866.028827 & $2\times1200$   & 0.38 & $151.2\pm8.6\pm4.0$  & $165.5\pm15.5$  & $159.9\pm8.5$  &  $75.6\pm8.5$   & -1.5 \\
RR1  & 55926.934245 & $1\times1800^*$ & 0.59 & $170.0\pm5.7\pm2.5$  & $159.2\pm13.6$  &                &                 & -1.7 \\
RR1  & 55952.995976 & $1\times1800^*$ & 0.17 & $115.8\pm4.8\pm3.2$  & $155.2\pm15.4$  &                &                 & -1.7 \\
RR2  & 55865.996308 & $2\times1200$   & 0.18 & $117.8\pm6.8\pm5.7$  & $156.0\pm16.7$  & $156.0\pm16.7$ &  $69.1\pm16.7$  & -0.9 \\
RR3  & 55917.913149 & $2\times1800$   & 0.42 & $155.0\pm4.2\pm6.9$  & $164.6\pm14.6$  & $157.0\pm9.8$  &  $79.9\pm9.8$   & -1.5 \\
RR3  & 55953.024432 & $1\times1800^*$ & 0.60 & $162.8\pm4.0\pm3.2$  & $150.9\pm13.2$  &                &                 & -2.0 \\
RR4  & 55916.996225 & $2\times1800$   & 0.49 & $181.5\pm9.8\pm10.9$ & $182.7\pm18.9$  & $182.7\pm18.9$ &  $93.9\pm18.9$  & -1.4 \\
\hline
RR5  & 55895.934069 & $2\times1200$   & 0.84 & $129.5\pm7.1\pm16.8$ & $98.2\pm23.3$   & $91.3\pm12.4$  &  $6.5\pm12.4$   & -2.6 \\
RR5  & 55922.944423 & $2\times1800$   & 0.11 &  $47.9\pm4.9\pm7.7$  & $94.5\pm17.7$   &                &                 & -2.4 \\
RR5  & 55926.018096 & $2\times1800$   & 0.15 &  $34.2\pm7.5\pm19.9$ & $76.0\pm25.8$   &                &                 & -2.5 \\
RR6  & 55926.967532 & $1\times1800^*$ & 0.60 & $117.3\pm5.4\pm2.5$  & $105.4\pm13.6$  & $105.4\pm13.6$ &  $18.1\pm13.6$  & -2.1 \\
RR7  & 55917.864566 & $2\times1800$   & 0.39 &  $87.5\pm3.9\pm9.9$  & $100.7\pm16.2$  & $100.7\pm16.2$ &  $23.0\pm16.2$  & -2.2 \\
RR8  & 55888.997603 & $2\times1800$   & 0.39 &  $88.1\pm8.2\pm2.2$  & $101.3\pm14.9$  & $101.3\pm14.9$ &  $22.4\pm14.9$  & -1.2 \\
\hline
RR9  & 56037.732088 & $2\times1800$   & 0.50 & $109.8\pm16.8\pm3.2$ & $109.8\pm17.1$  & $109.8\pm20.9$ &  $38.0\pm20.9$  & -2.0 \\
RR10 & 56037.175892 & $2\times1800$   & 0.44 & $152.1\pm11.8\pm5.7$ & $159.3\pm13.1$  & $159.3\pm17.8$ &  $88.6\pm17.8$  & -1.9 \\
\hline
AV Peg & 55917.604262 & $1\times270$    & 0.41 & $-75.5\pm7.8\pm11.9$ & $-64.8\pm14.2$ & $-64.8\pm14.2$ & $135.1\pm14.2$ & 0.1
\enddata
\tablenotetext{a}{Short name.}
\tablenotetext{b}{Reduced Heliocentric Julian Date when the spectrum was taken (HJD - 2400000).}
\tablenotetext{c}{Exposure times (* denotes Keck-I/LRIS setup).}
\tablenotetext{d}{Pulsation phase when the spectrum was taken.}
\tablenotetext{e}{Radial velocity (not corrected for pulsations) and its uncertainties (cross-correlation and zero-point).}
\tablenotetext{f}{Systemic (center-of-mass) velocity (corrected for pulsations) and its uncertainties (cross-correlation, zero-point, and model errors added in quadrature).}
\tablenotetext{g}{Weighted average of multiple heliocentric systemic velocities and its uncertainty.}
\tablenotetext{h}{Galactocentric rest-frame velocity and its uncertainty.}
\tablenotetext{i}{Metallicity.}
\end{deluxetable}

\clearpage

\begin{deluxetable}{lcl}
\rotate
\tabletypesize{\scriptsize}
\tablecolumns{3}
\tablewidth{0pc}
\tablecaption{ELODIE Template Stars\label{elodie_templates}}
\tablehead{
\colhead{Name} & \colhead{ELODIE ID Number} & \colhead{Spectral Type}
}
\startdata
HD 49933  & 00282 & F2V  \\
HD 140283 & 00480 & sdF3 \\
HD 102870 & 00414 & F8V  \\
HD 693    & 00004 & F5V  \\
HD 7476   & 00690 & F5V  \\
HD 13555  & 00998 & F5V  \\
HD 3268   & 00010 & F7V  \\
HD 222368 & 00855 & F7V  \\
HD 19994  & 00084 & F8V  \\
HD 22484  & 00092 & F9V  \\
HD 22879  & 00096 & F9V
\enddata
\end{deluxetable}

\clearpage

\begin{deluxetable}{lrrrr}
\rotate
\tabletypesize{\scriptsize}
\tablecolumns{3}
\tablewidth{0pc}
\tablecaption{Equivalenth-width Standard Stars\label{EW_stds}}
\tablehead{
\colhead{Name} & \colhead{${\rm W^\prime(K)}$} & \colhead{${\rm W^\prime(H\delta)}$} & \colhead{${\rm W^\prime(H\gamma)}$} &
\colhead{${\rm W^\prime(H\beta)}$} \\
\colhead{} & \colhead{${\rm mean\pm sd}$} & \colhead{${\rm mean\pm sd}$} &
\colhead{${\rm mean\pm sd}$} & \colhead{${\rm mean\pm sd}$}
}
\startdata
Kopff 27   & $1.72\pm0.06$ & $10.5\pm0.4$   & $10.2\pm0.3$  & $8.84\pm0.05$ \\
HD 180482  & $2.42\pm0.07$ & $10.26\pm0.09$ & $9.2\pm0.1$   & $8.7\pm0.1$   \\
BD+25 1981 & $2.79\pm0.05$ & $5.0\pm0.2$    & $5.4\pm0.3$   & $5.12\pm0.03$ \\
HD 155967  & $6.6\pm0.2$   & $3.0\pm0.3$    & $3.89\pm0.08$ & $3.95\pm0.09$ \\
HD 112299  & $8.4\pm0.2$   & $1.9\pm0.2$    & $2.10\pm0.07$ & $3.02\pm0.06$ \\
\hline
Kopff 27   & 1.89 & 9.68 & 10.44 & 8.89 \\
BD+25 1981 & 2.90 & 5.47 & 6.05  & 5.08 \\
HD 155967  & 6.90 & 2.75 & 3.84  & 4.05 \\
HD 112299  & 8.97 & 1.64 & 2.10  & 2.98
\enddata
\tablecomments{The horizontal line separates equivalent-width standard stars
observed with DBSP (over three nights) and LRIS (over a single night).}
\end{deluxetable}

\clearpage

\clearpage

\clearpage

\clearpage

\end{document}